\documentclass[sigconf,screen,nonacm]{acmart}

\AtBeginDocument{%
  \providecommand\BibTeX{{%
    \normalfont B\kern-0.5em{\scshape i\kern-0.25em b}\kern-0.8em\TeX}}}

\copyrightyear{2024}
\acmYear{2024}
\setcopyright{rightsretained}
\acmConference[PDC '24 Vol. 2]{Participatory Design Conference 2024, Vol. 2: Exploratory Papers and Workshops}{August 11--16, 2024}{Sibu, Malaysia}
\acmBooktitle{Participatory Design Conference 2024, Vol. 2: Exploratory Papers and Workshops (PDC '24 Vol. 2), August 11--16, 2024, Sibu, Malaysia}
\acmDOI{10.1145/3661455.3669890}
\acmISBN{979-8-4007-0654-7/24/08}

\begin{document}

\title{Envisioning Collaborative Futures: Advancing the Frontiers of Embedded Research}

\author{Anna R. L. Carter}
\author{Kyle Montague}
\author{Reem Talhouk}
\affiliation{%
 \institution{Northumbria University}
 \city{Newcastle}
 \country{UK}}

\author{Ana O. Henriques}
\author{Hugo Nicolau}
\affiliation{%
 \institution{ITI/LARSyS, Instituto Superior Técnico, Universidade de Lisboa}
 \city{Lisbon}
 \country{Portugal}}

\author{Tiffany Knearem}
\affiliation{%
 \institution{Google}
 \city{Boston}
 \country{USA}}

\author{Ceylan Besevli}
\email{c.besevli@ucl.ac.uk}
\affiliation{
    \institution{University College London}
    \city{London}
    \country{UK}
}

\author{Firaz Peer}
\email{firazpeer@uky.edu}
\affiliation{
    \institution{University of Kentucky}
    \city{Kentucky}
    \country{USA}
} 

\author{Clara Crivellaro}
\email{Clara.Crivellaro@newcastle.ac.uk}
\affiliation{
    \institution{Open Lab, Newcastle University}
    \city{Newcastle}
    \country{UK}
} 

\author{Sarah Rüller}
\email{sarah.rueller@uni-siegen.de}
\affiliation{%
 \institution{University of Siegen}
 \city{Siegen}
 \country{Germany}}

%Possibles to ask:
%%Firaz Peer, Clara Crivellero

\renewcommand{\shortauthors}{Carter et al.}

%% The abstract is a short summary of the work to be presented in the article.
\begin{abstract}
Participatory design initiatives, especially within the realm of digital civics, are often integrated and co-developed with the very citizens and communities they intend to assist. Digital civics research aims to create positive social change using a variety of digital technologies. These research projects commonly adopt various embedded processes, such as commissioning models \cite{dcitizensproj22}. Despite the adoption of this process within a range of domains, there isn't currently a framework for best practices and accountability procedures to ensure we engage with citizens ethically and ensure the sustainability of our projects. This workshop aims to provide a space to start collaboratively constructing a dynamic framework of best practices, laying the groundwork for the future of sustainable embedded research processes. The overarching goal is to foster discussions and share insights that contribute to developing effective practices, ensuring the longevity and impact of participatory digital civics projects.
\end{abstract}




%% Keywords. The author(s) should pick words that accurately describe the work being presented. Separate the keywords with commas.

%\centering
 % \includegraphics[width=\textwidth]{}
  %\caption{Visible Caption - this is if you want a picture before the abstract}
  %\Description{Alt text}
  %\label{fig:teaser}

\maketitle

\section{Motivation}
In the expanding landscape of participatory design, there is a common acknowledgement of the pivotal role played by citizens, especially in the co-creation of community-specific technologies and beyond. This recognition extends across diverse domains such as Participatory Design \cite{carter22lookout, erling10pddemoc}, Digital Civics \cite{pires20educatorprog, neto21robot, peer23refugees, carter24sig}, Neurodivergence and HCI \cite{piedade23pdkit, spiel22adhd}, Activism \cite{bardzell11sex, michie18her}, and Feminist HCI \cite{bardzell10feminist, henriques23feministethics}, highlighting the profound impact of citizen involvement beyond traditional project boundaries. In this context, there's a lack of guidance on incorporating citizens into projects ethically and sustainably. Academic timelines, driven by grants and paper deadlines, often clash with the unpredictable timelines of the communities we engage with. This misalignment raises questions about mitigating risks, planning exit strategies, and managing expectations from the project's outset. The challenge is ensuring that our projects, designed within research time constraints, align seamlessly with the ongoing and often prolonged challenges faced by the charities or community groups we collaborate with. The ultimate goal is to leave the community with the ability to sustainably continue the initiatives without ongoing external support, if that is not something that can continue to be available.

Moving beyond the simple inclusion of sustainability in initial planning, our objective is to explore the possible connections between citizen engagement and the development of sustainable projects. Ensuring the sustainability of participatory design projects is crucial not only during the initial stages but also throughout the entire project life-cycle to guarantee long-term success and real impact on communities \cite{poderi18sustainablepd}. The ultimate goal should be for communities to become fully independent, no longer needing facilitators in these ventures. Through sustained, long-term collaboration, communities can develop a sense of ownership in the projects, leading to the continued relevance and effectiveness of the digital solutions and refocusing the work on the communities themselves \cite{harrington19equitable}. This exploratory investigation will not only enhance our understanding of embedded research but also encourages us to contemplate inventive solutions and ethical frameworks that go beyond the traditional confines of technology research. Therefore, within this workshop, we aim to navigate between the intersection of citizen involvement and the sustainability of embedded project development.

\section{Goals and Questions}
The main goal of the workshop is to bring together researchers, industry members, practitioners and interested citizens to envision the future of embedded research. The overarching aim is to strengthen the community of researchers and build a network for future collaboration. Emphasising ethical and sustainable principles, the workshop will focus on establishing the groundwork for a robust framework to uphold collaborative efforts within the community in embedded research. Our intention is to further develop this framework beyond the workshop, engaging the community in its ongoing construction.

The workshop will explore the current practices used through an entire project lifecycle, including how to consider exit strategies from project initiation, thinking about community benefits and strength from the beginning \cite{costanza20justice}. These explorations will be grounded in real-world project examples, such as the use of commissioning models, which are increasingly popular as a tool for community-based projects \cite{johnson21commissioning, dcitizensproj22}. To guide this exploration, the following questions arise:

\begin{itemize}
    \item How can a suitable framework/toolkit be created to support researchers, industry members and citizens alike across participatory design domains?
    \item What are the most pressing challenges faced within embedded research, and how can we mitigate them? 
\end{itemize}

\section{Workshop Structure}
The workshop will be a full-day, hybrid event that will combine positional presentations, practical, hands-on activities, and discussion to support innovation and networking. The hybrid format will ensure we maximise attendance and engagement both prior to, during and after the conference. We aim to assign an in-person laptop member to each table to enable online members to be part of the in-person groups with break out rooms assigned to each table. A collaborative whiteboard tool, or another collaborative online whiteboard will be used to upload collaborative images in person and remotely. This whiteboard environment will be designed and made available to participants prior to the event to enable community building. 

\subsection{Pre-Workshop Activities (Asynchronous)}
\emph{\textbf{Introduce Yourself:}} Participants will be invited to a discord channel where they can introduce themselves and interact before the workshop. 

\subsection{Synchronous Hybrid Event}
Indicative format and activities, breaks and lunch will be confirmed as per the PDC 2024 schedule:
\\
\emph{\textbf{Introduction \& Welcome:}} Organiser introductions and scoping the problem. (10 minutes)
\\
\emph{\textbf{Participant Pecha Kucha:}} Each participant will have up to $2$ minutes to present their research domain and interest in the workshop topic. This ensures a comprehensive understanding of the diverse backgrounds and perspectives of participants. (Up to 35 minutes)
\\
\emph{\textbf{Commissioning a project}} This activity will focus on exploring effective strategies for commissioning projects that promote participatory design research across a variety of global contexts. Participants will be tasked with brainstorming the optimal process for co-creating initiatives and projects, where participants actively contribute to shaping agendas and making decisions about service provisions. (45 minutes) 
\\
\emph{\textbf{Break:}} Tea/coffee break (30 minutes)
\\
\emph{\textbf{Navigating Embedded Research Challenges:}} Throughout each embedded research process, challenges will present themselves that require careful management and are often unexpected. For example, a disagreement in value difference, having to put our own biases aside to work with organisations with different core values or dealing with unexpected outcomes that can lead to trauma. Within this activity, participants will be invited to bring their own experienced challenges for prompting and receive predefined challenges to discuss. (45 minutes)
\\
\emph{\textbf{Sustainable Embedded Research:}} A crucial challenge in embedded research is navigating project exits and leaving projects with the capacity to continue. This activity will discuss the importance of considering the end right from the project's inception. It aims to foster discussions on implementing continuous checks and a sustainable narrative throughout projects. It will prompt participants to detail their ideas for sustainable exit plans and establish and manage expectations from the outset. (45 minutes)
\\
\emph{\textbf{Break:}} Lunch break (60 minutes)
\\
\emph{\textbf{Participatory Design Activity:}} This activity will focus on brainstorming the optimal communication process for communicating all the goals and limitations identified throughout the previous activities, applying them to different embedded research contexts (one per group). Participants will be grouped and provided with storyboards illustrating interactions with fictional NGO's. Their task will be to annotate these storyboards using post-it notes and sketches, focusing on the most effective communication processes to mitigate the challenges faced with an aim to contribute to the sustainable advancement of the organisations. Each group will share their suggestions, serving as the foundational prompts for discussing broader implications and best practices in embedded research throughout the remainder of the workshop. (60 minutes)
\\
\emph{\textbf{Speculating the future of Embedded Research:}} This will be a speculative exercise where we use prompts that provoke our understanding of embedded research and the frameworks/tools we can use in the future to aid sustainability without being tied down by technology. The work will imagine alternatives to our current approaches and discuss how embedded research would be envisioned in such a scenario. The goal would be to reconsider our future vision and perhaps come up with novel interfaces and usage areas. (60 minutes)
\\
\emph{\textbf{Break:}} Tea/coffee break (30 minutes)
\\
\emph{\textbf{Moving Forward:}} Bringing together the previous discussions, groups will discuss the three key considerations they believe could shape the future of embedded research within digital civics to create more ethical and sustainable interventions across the next decade. They will write these onto three separate post-it notes and present them back to the group. The participants will then place these onto the wall and start to rate them from most to least important to incorporate into their own work. This exercise will help to build the framework for the future of sustainable embedded research. (25 minutes)
\\
\emph{\textbf{Reflections \& Wrap-Up:}} End of day thoughts and takeaways.
\\
\emph{\textbf{Dinner and Further Discussions:}} Participants will be invited to continue discussions at the conference welcoming reception.

\subsection{Post-Workshop Activities (Asynchronous)}
\emph{\textbf{Reflections on the event:}} We will invite thoughts, ideas and discussion points to keep the conversation going. There will be an area of the online whiteboard to place these items. In addition, an invitation to join the DCitizens Discord and Seminar Group (our ongoing \href{dcitizens.eu}{EU project}) to expand the network will be provided.

\section{Planned Outcomes}
This workshop will build upon the network forming within the Discord group and seminar series already in motion and serve as an extension, not only expanding the community but also with aspirations to set up a SIGCHI chapter. Additionally, we plan to write up an Interactions article based on this discussion and organise follow-on workshop's at conferences, including CHI $2025$ and PDC $2026$. This initiative seeks to enhance collaboration and foster knowledge exchange within the digital civics community and eventually set up a framework for the future of embedded research. Our goal is to kickstart a continuous dialogue, expressing varied perspectives, extracting insights, and laying the groundwork for collaboratively creating a guiding framework that future participatory designers can draw upon when navigating the complexities of this domain.

\section{Intended Audience and Recruitment}
We will gather $6$-$30$ participants whose research interests align with the goals and questions we have outlined. Building upon the foundations laid by our previous work \cite{carter22lookout, talhouk22dial,talhouk20food,Puussaar22sense, knearem24covid, knearem21food}, as well as the ongoing Horizon project \cite{dcitizensproj22} 
there is a network of researchers, practitioners, community members, and policymakers who share an interest in the human aspects of digital civics that we can reach out to. Additionally, many of the organisers will promote on social media and can utilise their networks to advertise the call, as well as using more traditional methods such as mailing lists. We will also direct people to our \href{https://dcitizens.eu}{website}, which we will keep regularly updated with useful information in the run up to the live conference dates. As the workshop is intended to be run in a hybrid fashion, we anticipate being able to offer a range of accessibility options (e.g. wheelchair access, captioning as available via the conference platform), and will make a "kit-list" for remote participants who wish to actively take part in the hands-on activities. 

\subsection{Call for Participation}
In early May 2024, we will distribute instructions asking participants to prepare a short statement, or pictorial (maximum 1 page) that includes a short description of their relevant professional and/or research practice in a maximum of 200 words and a motivation statement related to their interest in the workshop's motivations, in a maximum of 500 words. Participatory design is not simply about text, so we will invite participants to use the formats creatively and welcome submissions in alternative formats. Participants should submit these prior to the early-bird deadline, with deadlines to be uploaded to the website. 

We are hosting a workshop aiming to support the design and development of the future of embedded research. The full-day workshop will combine presentations, hands-on practical experiences (in person and online) and group discussions. All accepted papers will be hosted and maintained on the DCitizens website. Following PDC Guidelines, all participants must register for both the workshop and for at least one day of the conference.
\\
\emph{\textbf{Paper/Pictorial submissions:}} Email submissions to Anna Carter: anna.r.l.carter@northumbria.ac.uk \\
\emph{\textbf{Workshop website: \url{https://dcitizens.eu/envisioning-collaborative-futures-pdc-2024/}}}

\section{Why PDC 2024?}
This workshop aligns with the conference theme, \emph{Connecting beyond participation: reaching out} because we want to discuss with the wider digital civics community the steps we can take to sustainably complete embedded research and ensure we are working with participants beyond initial participation. We want to involve them in the entire process from commissioning the projects to participating to working collaboratively in order to provide them with the tools to sustain these projects when we move forward. The legacy of the conference's commitment to democratic practices makes it the perfect place to bring together researchers and industry members from varied fields working on digital civics to lay the groundwork for the future of embedded research.

\section{Organisers}
The workshop aims to be an inclusive event that will enable the cross-pollination of various techniques and issues raised within the completion of embedded research. The workshop has been designed to cater to attendees with varying backgrounds and focus, ensuring that the discourse is accessible and relevant to all. The organisers are from diverse institutions, industries, and disciplines, with extensive experience in designing, studying and publishing on open participatory processes and digital civics spanning a wide range of geographical, institutional and social contexts. Therefore, we are well-equipped to accommodate and facilitate discussions across the theme of embedded research, having completed it within a range of contexts and therefore, fostering an environment where insights from different backgrounds converge and enrich the conversation. This workshop aims to brainstorm ideas and build an interdisciplinary community to continue the conversation around sustainable embedded research within digital civics.

\begin{acks}
We thank our funding bodies at the European Commission (101079116 Fostering Digital Civics Research and Innovation in Lisbon), EPSRC (EP/T022582/1 Centre for Digital Citizens - Next Stage Digital Economy Centre) and the Portuguese Recovery and Resilience Program (PRR), IAPMEI/ANI/FCT under Agenda C645022399-00000057 (eGamesLab).
\end{acks}

%% The next two lines define the bibliography style to be used, and the bibliography file.
\bibliographystyle{ACM-Reference-Format}
\bibliography{dcitcom}

\end{document}